\documentclass[preprint,8pt,5p,twocolumn]{elsarticle}

\usepackage{epsfig}

\usepackage{amssymb}
\usepackage{amsmath}

\usepackage[nottoc]{tocbibind}

\journal{Journal of Colloid and Interface Science}

\begin{document}

\begin{frontmatter}


\title{Early blood–material interfacial events and capillary transport on nanoparticle‑modified nanofibers \tnoteref{label1}}

\title{} 


\author[a,b]{Romain Scarabelli\fnref{eq}}
\author[c]{Mehdi Abbasi\fnref{eq}}
\author[d]{Magali Gary-Bobo}
\author[b]{Christophe Drouet}
\author[c]{Marc Leonetti\corref{cor1}}
\ead[cor1]{marc.leonetti@univ-amu.fr}
\cortext[cor1]{Corresponding author}

\author[a]{Ahmed Al-Kattan\corref{cor2}}
\ead[cor2]{ahmed.al-kattan@univ-amu.fr}
\cortext[cor2]{Corresponding author}

\fntext[eq]{RS and MA contributed equally to this work.}

\affiliation[a]{organization={Aix Marseille Univ, CNRS, LP3},
            city={Marseille},
            country={France}}

\affiliation[b]{organization={CIRIMAT, Université de Toulouse/INP/CNRS, 4 allée Emile Monso},
            city={Toulouse},
            postcode={31030},
            country={France}}

\affiliation[c]{organization={Aix Marseille Univ, CNRS, CINaM, Turing Centre for Living Systems},         
            city={Marseille},
            country={France}}

\affiliation[d]{organization={IBMM, Université de Montpellier, CNRS, ENSCM, 1919 Route de Mende},
            city={Montpellier},
            postcode={34293},
            country={France}}

\author{} 

\begin{abstract}
Electrospun poly($\epsilon$‑caprolactone) (PCL)nanofibrous mats are widely considered for blood-contacting wound dressings and small-diameter vascular applications; however, their intrinsic hydrophobicity limits rapid wetting and controlled interaction with blood. In this work, we modulate the interfacial response of PCL nanofibers by incorporating oxide-shelled silicon nanoparticles (SiNPs) synthesized by pulsed laser ablation in liquid, a ligand-free approach that avoids organic stabilizers and preserves surface reactivity. Two composite architectures were designed: SiNPs embedded within the fiber bulk (PAC-1, -4, -16) and SiNPs preferentially exposed at the fiber surface (SPAC-1, -4, -16), with systematically increasing nanoparticle loadings. Structural characterization confirmed the retention of a homogeneous fibrous morphology and the targeted nanoparticle distribution. The dynamic interaction with whole blood was quantified using time-resolved contact-angle measurements, complemented by top-view optical microscopy and three-dimensional profilometry of dried droplets. Pristine PCL remained strongly hydrophobic, exhibiting a high apparent contact angle that decreased only marginally over time ($\approx$110° to $\approx$100° over 20 min), whereas a hydrophilic PCL functionalized with APTES showed rapid spreading. Incorporation of SiNPs within the fiber volume led to only a moderate enhancement of wettability (final angles $\approx$80–90°), and dried droplets retained compact morphologies with limited spreading. In contrast, surface-decorated mats displayed a sharp, concentration-dependent transition toward highly wettable behavior: for SPAC-16, the contact angle fell below 20°, droplet profiles became markedly flattened, and microscopy revealed extended plasma-rich regions surrounding a red-cell-rich core, indicative of pronounced phase separation within the nanofibrous network. Consistently, gravimetric measurements showed substantial increases in both water uptake (from $\approx$400$\%$ for PCL to >700$\%$ for SPAC-16) and blood uptake (up to $\approx$1200$\%$ for PAC-16 and $\approx$1050$\%$ for SPAC-16).

Overall, these results establish laser-synthesized SiNPs as an effective and chemically simple strategy to control blood wetting, imbibition, and phase separation in electrospun PCL nanofibers. Importantly, nanoparticle localization—within the fiber bulk or at the fiber surface—governs distinct regimes of interfacial behavior. This work provides mechanistic insight into early blood–material interactions on nanofibrous substrates and offers clear design guidelines for tailoring fluid management and hemocompatibility in advanced wound dressing and blood-contacting biomaterials.
\end{abstract}




\end{frontmatter}



\section{Introduction}
\label{sec1}

Blood--material interactions are a critical consideration for any device or dressing intended for direct contact with blood. As soon as a material is exposed to blood, a complex sequence of interfacial events is triggered, including protein adsorption and conformational rearrangement, platelet adhesion and activation, fibrin formation, as well as complement and leukocyte recruitment. Together, these processes determine whether the interface supports physiological hemostasis or promotes pathological thrombosis and inflammation \cite{spijker2003influence,deppisch1998blood,brash2000exploiting}. In the context of wound dressings and vascular substitutes, this balance is particularly delicate: the material must enable rapid bleeding control and stable clot formation at the injury site, while simultaneously limiting excessive platelet activation or uncontrolled thrombus growth that could impair downstream perfusion. Achieving such controlled and predictable blood–material interactions therefore remains a central challenge in the design of hemocompatible materials \cite{newman2023challenge,ishihara2018blood,segal1998effects}.

Electrospun nanofibrous (NF) scaffolds have attracted considerable attention as platforms for advanced wound dressings and small-diameter vascular grafts, owing to their high specific surface area, interconnected porosity, and fibrillar architecture that closely resembles the native extracellular matrix \cite{wiegand2019hemostatic,nakielski2019blood,milleret2012influence,li2019superhydrophobic}. These structural features promote efficient fluid management, gas exchange, and cell infiltration, all of which are essential for effective wound healing. Among the polymers commonly used for electrospinning, poly($\epsilon$-caprolactone) (PCL) occupies a prominent position due to its biodegradability, mechanical robustness, and regulatory approval for several implantable medical devices \cite{bhullar2017design,cho2015preparation,kim2008electrospun,baker2016determining,azari2021electrospun}. Despite these advantages, pristine PCL is intrinsically hydrophobic and exhibits a low surface free energy. Such surface properties are known to promote non-physiological plasma protein adsorption and to induce platelet adhesion and activation patterns that are suboptimal for hemocompatibility \cite{zhou2020facile,leszczak2013hemocompatibility,leszczak2014improved,woodruff2010return}. In the context of wound dressings and intravascular constructs, this behavior can lead to delayed blood wetting, spatially heterogeneous clot formation, and, in certain cases, an elevated risk of thrombotic complications.

To overcome these limitations, numerous chemical and physical surface modification strategies have been investigated. These include plasma or corona treatments followed by the grafting of hydrophilic chains (e.g., PEG or zwitterionic brushes), blending PCL with natural polymers such as collagen, gelatin, or chitosan, immobilization of anticoagulant agents (e.g., heparin or hirudin), and layer-by-layer assembly of polyelectrolyte films \cite{recek2016cell,wang2012novel,zhu2002surface,chen2010surface,de2008biofunctionalization,decher2012multilayer,dash2012poly}. Although many of these approaches improve wettability and attenuate platelet adhesion under in vitro conditions, they often entail practical limitations, including limited long-term stability, multistep fabrication protocols, and the use of reactive chemicals or organic solvents that are difficult to fully eliminate. Moreover, most strategies primarily focus on surface chemistry and charge, while comparatively less attention has been paid to the role of nanoscale topography and capillary effects in highly porous fibrous architectures, and to how these features govern blood transport and phase separation during the earliest stages of contact \cite{ratner2000blood}.

Inorganic nanoscale fillers provide an alternative route to engineer surface energy, roughness and bioactivity without relying on complex chemistries. Silica‑based nanoparticles and bioactive glasses have been incorporated into polymeric matrices to increase hydrophilicity, modulate protein adsorption or confer osteoconductivity \cite{rezwan2006biodegradable, gaharwar2014nanocomposite}. Silicon nanostructures are of particular interest because their native oxide shell bears silanol groups that can engage in hydrogen bonding and specific interactions with proteins, while the underlying silicon core offers additional functionalities (e.g. photothermal, electronic or degradation‑derived ionic species) \cite{anglin2008porous}. Most studies to date, however, have focused on chemically synthesized silica or organosilane‑modified particles, which typically carry residual ligands or by‑products that may complicate biological interpretation. Furthermore, their use in the context of blood‑contacting electrospun scaffolds has been only sporadically examined, and systematic analyses of how nanoparticle loading and spatial distribution within nanofibrous mats affect blood wetting, imbibition and clot architecture are still scarce \cite{mamaeva2013mesoporous, abrigo2014electrospun, gaharwar2014nanocomposite}.

A related gap concerns the level at which blood–material interactions are usually characterized. Conventional hemocompatibility evaluation tends to emphasise end‑point assays—hemolysis, coagulation and complement activation times, platelet adhesion counts—performed after relatively long exposures \cite{bernard2018biocompatibility}. While these metrics are indispensable, they provide limited insight into the early‑time interfacial phenomena that precede them. In highly porous fibrous substrates, these initial events involve dynamic spreading and penetration of blood into the mat, capillary‑driven phase separation between cellular and plasma components, and spatially heterogeneous protein deposition. The interplay between surface chemistry, fiber morphology and these non‑equilibrium wetting processes is only partially understood, yet it is precisely during these first seconds to minutes that the structure of the provisional matrix is established and the subsequent biological response is largely predetermined.

Laser‑synthesized silicon nanoparticles (SiNPs) obtained by pulsed laser ablation of a silicon target in liquid offer an attractive tool to address some of these issues. This physical synthesis route yields nearly ligand‑free, oxide‑shelled SiNPs with a relatively narrow size distribution and high colloidal purity, avoiding the surfactants, reducing agents or organosilane precursors common to wet‑chemical methods \cite{zhang2017laser, kabashin2016theranostic, al2016ultrapure}. Such particles can therefore be introduced into or onto polymeric matrices without adding poorly defined organic species that might themselves influence blood responses. In addition, SiNPs provide a modular handle to tune surface polarity and roughness at the nanometre scale through simple variations in loading and localization, without altering the PCL backbone or requiring post‑spinning reactions.

To our knowledge, the combination of electrospun PCL nanofibers with laser‑synthesized SiNPs has not yet been explored in a hemocompatibility context, and several fundamental questions remain open. How does the presence of oxide‑shelled SiNPs affect the wetting dynamics of whole blood on PCL nanofibrous mats? Does embedding the particles within the fibre volume have the same effect as decorating the fiber surface, or do these two configurations lead to distinct regimes of droplet spreading, imbibition and blood phase separation? To what extent can nanoparticle‑induced changes in surface free energy be correlated with macroscopic fluid absorption and with the morphology of the blood deposit at the fiber scale? Answering these questions would not only clarify the role of SiNPs as interfacial modifiers, but also provide general design guidelines for nanofibrous dressings aiming at controlled hemostasis and improved hemocompatibility.

In this work we address these issues by systematically investigating blood interactions with electrospun PCL nanofibers functionalized with laser‑synthesized SiNPs. We prepare two families of mats in which the same SiNPs are either incorporated throughout the PCL fibre volume (PAC series) or preferentially exposed at the fibre surface (SPAC series), each at three different nanoparticle loading. After physicochemical and morphological characterization of the fibers, we analyze in detail the interaction of whole citrated blood with these substrates. Dynamic wetting is quantified through time‑resolved contact‑angle measurements, complemented by digital microscopy and 3D optical profilometry to reconstruct the final droplet shape and to visualize blood spreading and phase separation on and within the fibrous network. Macroscopic water and blood uptake are then measured to evaluate the ability of the mats to manage fluids under immersion conditions. Finally, these interfacial and transport properties are considered together with basic hemocompatibility endpoints to assess how SiNP functionalization modulates the early events that govern blood–material compatibility.

By combining a ligand‑free nanoinorganic modifier with a clinically relevant polymer and by placing emphasis on the multiscale description of blood wetting and imbibition, this study seeks to bridge the gap between classical surface chemistry approaches and the complex, dynamic behavior of blood in contact with porous fibrous substrates. The insights gained here should be useful not only for the rational design of PCL‑based wound dressings and vascular scaffolds, but more broadly for understanding how nanoscale surface modification strategies translate into macroscopic hemocompatible performance.
\section{Materials and methods}
\label{sec2}

\subsection{Materials and reagents}
\label{subsec2.1}
$\epsilon$-polycaprolactone (PCL) of 80000 Da average molecular weight and aminopropyltriethoxysilane (APTES) were supplied by Sigma Aldrich (Germany). Glacial acetic acid and absolute ethanol were ordered from Fisher chemicals (France). Ultrapure water was obtained using a MilliQ water purifier. Silicon wafers were ordered from Goodfellow (UK). Whole blood samples were obtained from the Etablissement Français du Sang (EFS) and used in accordance with institutional and ethical guidelines.

\subsection{Laser synthesis of silicon nanoparticles}
\label{subsec2.2}
Silicon nanoparticles production and their characterization was carried out following the previously published method \cite{scarabelli2026combination}. Si-NPs were produced by pulsing a femtosecond laser (Light Conversion Carbide CB1-05 Yb:KGW) on a Si wafer target immersed in ethanol. Laser parameters were fixed at a pulse duration of 400 fs, delivering 75.8 $\mu$J per pulse, with a repetition rate of 10 kHz at 1028 nm. The SiNPs were characterized with a high-resolution transmission electron microscope (HR-TEM, JEOL JEM 3010) working in imaging and diffraction modes.

Various concentrations of SiNPs were obtained by centrifugation of the initial SiNP suspension (C1). The concentration of the SiNPs was measured by inductively coupled plasma mass spectrometry (ICP-MS), resulting in 14 $\mu$g/mL for C1, 56 $\mu$g/mL for C4 and 224 $\mu$g/mL for C16. SiNPs hydrodynamic diameter in solution and zeta potential were assessed using a Zetasizer Nano ZS instrument (Malvern Instruments).

\subsection{XPS analysis}
\label{subsec2.3}
The nanoparticles oxidation state was determined using XPS measurements using a Thermo Scientific K-alpha instrument. Photoelectron emission spectra were recorded using $Al-K \alpha$ radiation ($h \nu = 1486.6 eV$) from a monochromatized source, with an X-ray spot size of 400 $\mu$m. The pass energy was set to 30 eV (0.1 eV step) for high-resolution scans and 100 eV (1 eV step) for survey scans. A flood gun was employed to neutralize charging effects. Measurements were done on freeze-dried SiNPs suspension after several cycles of vacuum pumping to remove trace moisture.

\subsection{Polymer preparation and functionalization with laser synthesized SiNPs}
\label{subsec2.4}
According to our previous work \cite{scarabelli2026combination}, PCL beads were dissolved in a 9:1 acetic acid/SiNP suspension with a total fraction of PCL of 17 $\%$ w/v. The solution was brought to 65 °C under vigorous stirring for 1 h until complete polymer dissolution. The functionalization of PCL with SiNPs has been achieved by adding APTES to the previously described polymer melt at a volume fraction of 1.7 $\%$, which has been optimized to ensure functionalization and electrospinnability.

\subsection{Electrospinning process and fiber characterization}
\label{subsec2.5}
The polymer melt was placed into a 5 mL Luer-lock plastic syringe which connected to the pump of the electrospinning device (Spinbox, Bioinicia Fluidnatek, Spain). Electrospinning parameters were set to 18.5 kV applied voltage, tip-to-collector distance of 14.25 cm and feeding rate of 0.005 mL/min. All experiments were conducted at room temperature. The obtained nanofibers were observed through SEM (Jeol JSM-6390) and HR-SEM (Jeol JSM-7900F) working at 25 and 10 kV respectively. NF width measurements were made using ImageJ software, measuring individual NF over multiple micrographs for each sample. For complementary analyses on the chemical nature of the fibres, FTIR spectra were recorded using a Bruker Vertex 70 spectrometer in the 80-4000 cm$^{-1}$ range, using 100 scans and a resolution of 4 cm$^{-1}$.

\subsection{Ultrasound-assisted dip-coating of SiNPs on the PCL NFs}
\label{subsec2.6}
Square samples of NFs were cut and immersed in a SiNP suspension before being placed in an ultrasonic bath for 5 minutes. NFs were then washed multiple times in deionized water to remove the unadhered NPs. The samples were then air-dried at room temperature.

\subsection{Pore size assessment}
\label{subsec2.7}
The porosity of the NFs samples was evaluated digitally from SEM micrographs using ImageJ, following a described procedure. To summarize, the pictures were converted in grayscale and their contrast and brightness was automatically adjusted. The images were processed using the software's native algorithm. A gaussian blur was then applied to homogenize the surface of the NFs. Thresholding 50$\%$ of the images allowed for the selection of the upper and intermediate layers of the NFs structure. Pore sizes were then measured on the binary images obtained.

\subsection{Contact angle measurement and droplet observations}
\label{subsec2.8}
\begin{figure}[htbp]
\centering
\includegraphics[scale=0.55]{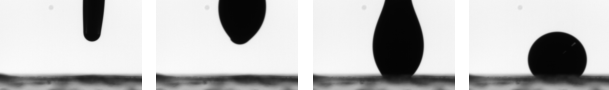}
\caption{Snapshot showing the deposition of a blood droplet onto the substrate using a calibrated micropipette during contact angle measurements.}\label{fig1}
\end{figure}

The contact angle of blood droplets on the different substrates was measured using an optical contact angle and contour analysis system (OCA 15EC, DataPhysics Instruments). Droplets were gently deposited onto the substrates surface using a calibrated pipette, as illustrated in Fig.\ref{fig1}. The droplet profile was recorded over time with a high-resolution camera, and the contact angle was determined by fitting the droplet contour using the SCA 20 software (DataPhysics Instruments). The droplet volume was fixed at 10µL in order to ensure good optical resolution of the droplet contour, minimize evaporation effects, and allow for reliable and reproductible contact angle determination while remaining below the capillary length to avoid significant gravitational deformation. After drying, the droplet were observed under a Keyence VK-X3000 laser microscope for surface rugosity and topological analysis, Keyence VH-X1 for high-definition images a Jeol JSM-6390 for scanning electron microscopy.

\subsection{Blood absorption test}
\label{subsec2.9}
Square samples of 2 $\times$ 2 cm of electrospun NFs were cut and placed in a hemolysis tube for 5 minutes and agitated by inversion. After removal, the unabsorbed surface blood was eliminated by lightly tapping absorbent lint-free paper on the samples. The electrospun mats mass was measured, and the blood absorption percentage was calculated using the following equation:
\begin{equation}
blood absorption (\%) = \frac{w_{wet} - w_{dry}}{w_{dry}} \times 100
\end{equation}
where $w_{wet}$ is the weight of the wet, bloodied sample and $w_{dry}$ is the weight of the dry sample, before the
experiment.
\subsection{Statistical analysis}
\label{subsec2.10}
All experiments were performed at least in triplicate. Data were statistically analyzed using one-factor ANOVA and the adequate multiple comparison tests. A P-value $ <$ 0.05 was considered as statistically significant.

\section{Results and discussion}
\label{sec3}

\subsection{Silicon nanoparticle synthesis and characterization}
\label{subsec3.1}

Synthesizing NPs using pulsed laser technology brings multiple advantages for the development of new medical devices. NPs produced via laser ablation are ultra-pure and free from both impurities and stabilizing molecules, thus reducing the risks of unwanted toxicity brought by residual solvents. The high purity of these NPs also simplifies their use as no purification steps are required, thus reducing waste production through these methods.

\begin{figure}[htbp]
\centering
\includegraphics[scale=0.45]{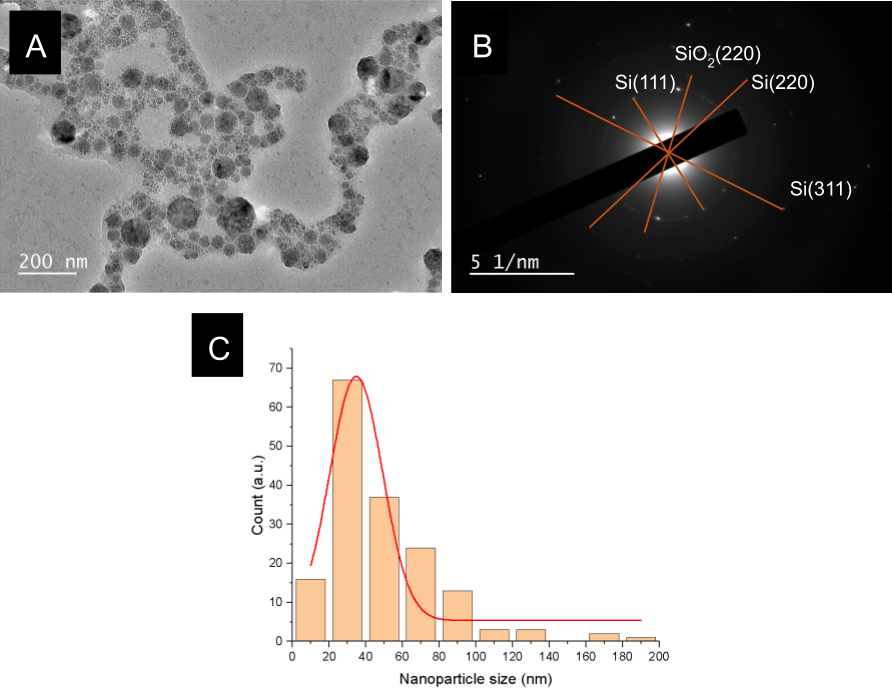}
\caption{Characterization of the silicon nanoparticles. (A): TEM micrograph of a sample ablated in ethanol. (B) Diffraction pattern of the SiNPS. (C) Nanoparticle size distribution}\label{fig2}
\end{figure}

We have previously demonstrated that SiNPs can play a crucial role in favoring cell proliferation both as a standalone proliferative agent or integrated within electrospun nanofibers \cite{murru2024assessment, scarabelli2026combination}. Following the protocol described in the experimental section, the colloidal suspensions of SiNPs were produced by means of a femtosecond laser. Laser ablation in ethanol allowed us to synthesize ultra-pure metallic SiNPs. The colloidal suspensions synthesized were firstly characterized using HR-TEM and X-ray diffraction Fig.\ref{fig2}. 

The synthesized SiNPs observed through TEM were of spherical shape with an average diameter of 35 $\pm$ 29 nm. Electron diffraction analysis allowed us to confirm the polycrystalline nature of the SiNPs, as the diffraction rings corresponding to the Si0 (111), (220), and (311) planes of the cubic structure of metallic Si were found in our samples, corresponding to the standard pattern provided by the ESRF (Fig.\ref{fig2}b). Additionally, the presence of the SiO2 (220) plane illustrated partial oxidation of the NPs, consistent with our previous findings. The zeta potential of the SiNPs was found to be negative, with a value of -43 $\pm$ 2 mV also confirming partial oxidation. Colloidal SiNP suspensions were concentrated by centrifugation, allowing us to obtain multiple suspensions of concentrations ranging from 14 (C1), 56 (C4) and 224 $\mu$g/mL (C16) as determined by ICPMS. 

\subsection{Nanofiber production}
\label{subsec3.2}

\begin{figure}[htbp]
\centering
\includegraphics[scale=0.5]{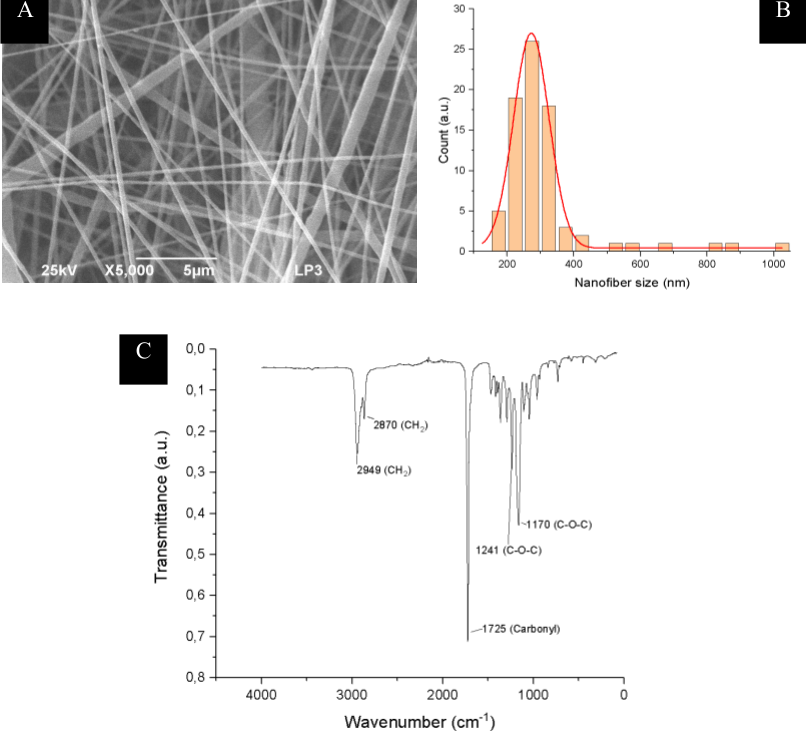}
\caption{Characterization of the pure PCL nanofibers. (A): SEM micrograph of the PCL NFs. (B): Size distribution graph. (C) FTIR spectrum.}\label{fig3}
\end{figure}

The above mentioned SiNPs were then integrated in an electrospun PCL matrix. All electrospinning parameters such as collector distance, applied voltage, feeding rate and electrospinning duration were previously optimized \cite{scarabelli2026combination}. This protocol, summarized in the experimental section, allowed us to minimize the number of defects such as polymer bead or spindle-like formation within the fibers. Briefly, the parameters were fixed as such: 14.25 cm tip to collector distance, 18.5 kV applied voltage, a feeding rate of 0.005 mL/min and 20 min electrospinning duration per batch.
After synthesis, the electrospun mats were characterized using SEM imagery and FTIR spectroscopy to ensure their structural and chemical integrity (Fig.\ref{fig3}). The PCL NFs width was measured using ImageJ at 272 $\pm$ 106 nm, remaining consistent with our previous work \cite{scarabelli2026combination}. The IR spectrum of the PCL mat also remained consistent with the literature, showing the characteristic peaks of PCL polymer: asymmetric CH2 stretching (2949 cm$^{-1}$), symmetric CH2 stretching (2870 cm$^{-1}$), carbonyl stretching (1725 cm$^{-1}$), asymmetric C-O-C stretching (1241 cm$^{-1}$) and symmetric C-O-C stretching (1170 cm$^{-1}$). These pure PCL mats were used as reference for the rest of this work, as PCL is biologically inert and completely biocompatible.

\begin{figure}[htbp]
\centering
\includegraphics[scale=0.4]{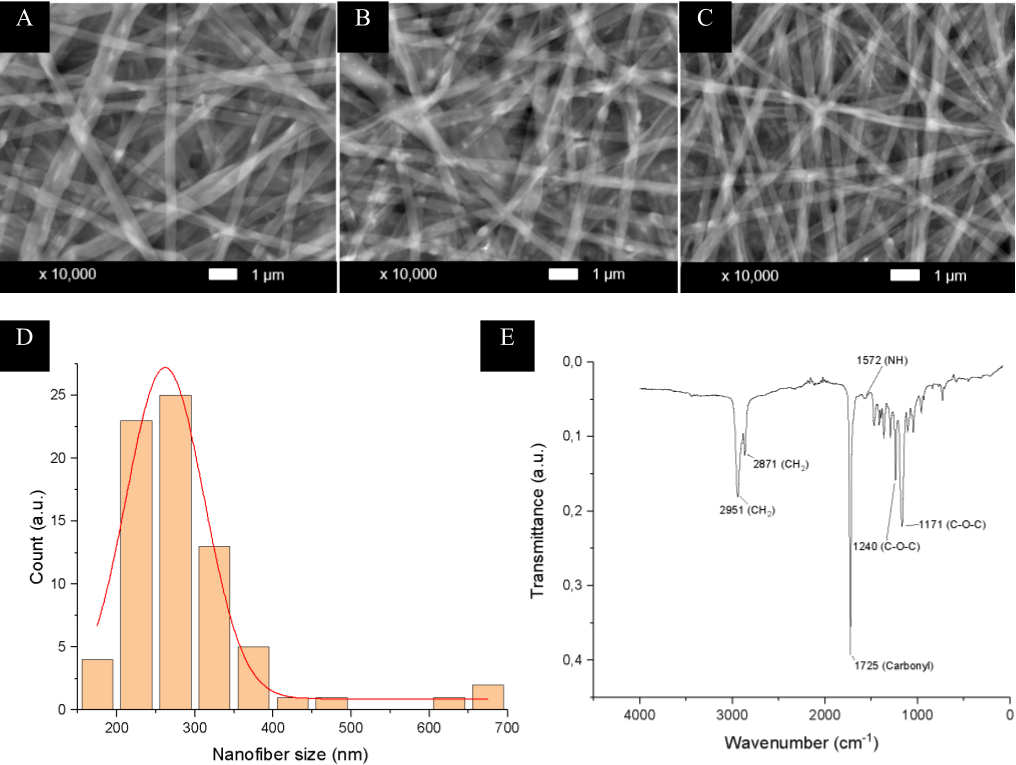}
\caption{Characterization of the NFs functionalized by SiNPs in the volume of the fibers. (A) HR-SEM micrograph in BED-C mode of a PAC-1 sample; (B) HR-SEM micrograph in BED-C mode of a PAC-4 sample; (C) HR-SEM micrograph in BED-C mode of a PAC-16 sample; (D) Average size dispersion of the SiNP-containing samples; (E) FTIR spectrum of a PAC-16 sample.}\label{fig4}
\end{figure}

To functionalize the PCL NFs, APTES, an aminosilane, was employed as a linker between the polymer and the SiNPs. Indeed, it was demonstrated \cite{scarabelli2026combination} that due to the hydrophobicity of PCL, unfavorable interactions prevent the inclusion of the dispersed SiNPs when introduced in the polymer melt. When added without the use of APTES, the SiNPs disrupted the homogeneity of the melt, causing the production of beaded NFs and exclusion of SiNPs from the polymer matrix. To facilitate the introduction of SiNPs in the NFs, APTES was introduced to the polymer melt before the addition of the SiNPs. Aminosilanes are widely used in literature to functionalize the surface of silicon oxide materials and NPs through reaction with the silane end group \cite{mahtabani2020silica, estevao2021mesoporous, rodio2016direct}. Additionally, the amine can favor hydrogen or covalent interactions with the polymer chain, thus leading to increased dispersion and compatibility between the SiNPs and the PCL NFs. Thus, using this strategy, three different concentrations of SiNPs suspensions were integrated into the volume of the NFs, replacing the deionized water, alongside APTES which was added in a volume fraction of 1.7 $\%$, named there after PAC-1, PAC-4 and PAC-16 according to their relative SiNP concentration. Formulations containing only PCL and APTES (without SiNPs) were also synthesized and named PA in the subsequent sections. All NFs obtained using this method were characterized using SEM and FTIR. HR SEM allowed us to observe in detail the integrated SiNPs in the volume of the NFs, as illustrated by (Fig.\ref{fig4}A-C). When observed in  backscattered electron imaging, the scintillating nano-objects were found to be dispersed along the whole fibrous structure, further confirming the favorable interactions taking place between the PCL and the SiNPs through APTES functionnalization. 
Interestingly, the NF sizes reported were smaller overall, with an average of 262 nm $\pm$100 nm. This phenomenon could be the result of the drop in viscosity induced by the APTES, as was reported in our previous work. Indeed, polymer melt viscosity is closely tied to the NF size during the electrospinning process \cite{stepanyan2016nanofiber}.
FTIR spectra remained in line with the pure PCL spectra, with the addition of the NH vibration at 1572 cm$^{-1}$ following the addition of APTES in the polymer melt. 
These formulations were described in a previous study as biocompatible, and a proliferative effect of the SiNPs, when present in the volume of the NFs, was determined.

\subsection{Ultrasonic dispersion of SiNPs on the electrospun mats}
\label{subsec3.3}

The previous NF formulations functionalized with SiNPs, although biocompatible and presenting advantageous proliferative properties, have their bioactive charge enclosed within the volume of the individual fibers. To adequately study the effect of these NPs on blood and blood components, it is critical that the interfacial interactions between SiNPs and blood through direct contact are also evaluated. Thus, an additional method of functionalization was carried out to decorate the surface of the PCL NFs with SiNPs. The most efficient way to achieve this goal was determined to be via dip-coating the NF mats. To ensure sufficient agitation and dispersion of the colloidal SiNPs throughout the surface of the NFs, this process was assisted by ultrasonication. To this end, pure PCL and PA mats were immersed in the various SiNPs suspensions and placed in an ultrasonic bath for 10 minutes. 

\begin{figure}[htbp]
\centering
\includegraphics[scale=0.32]{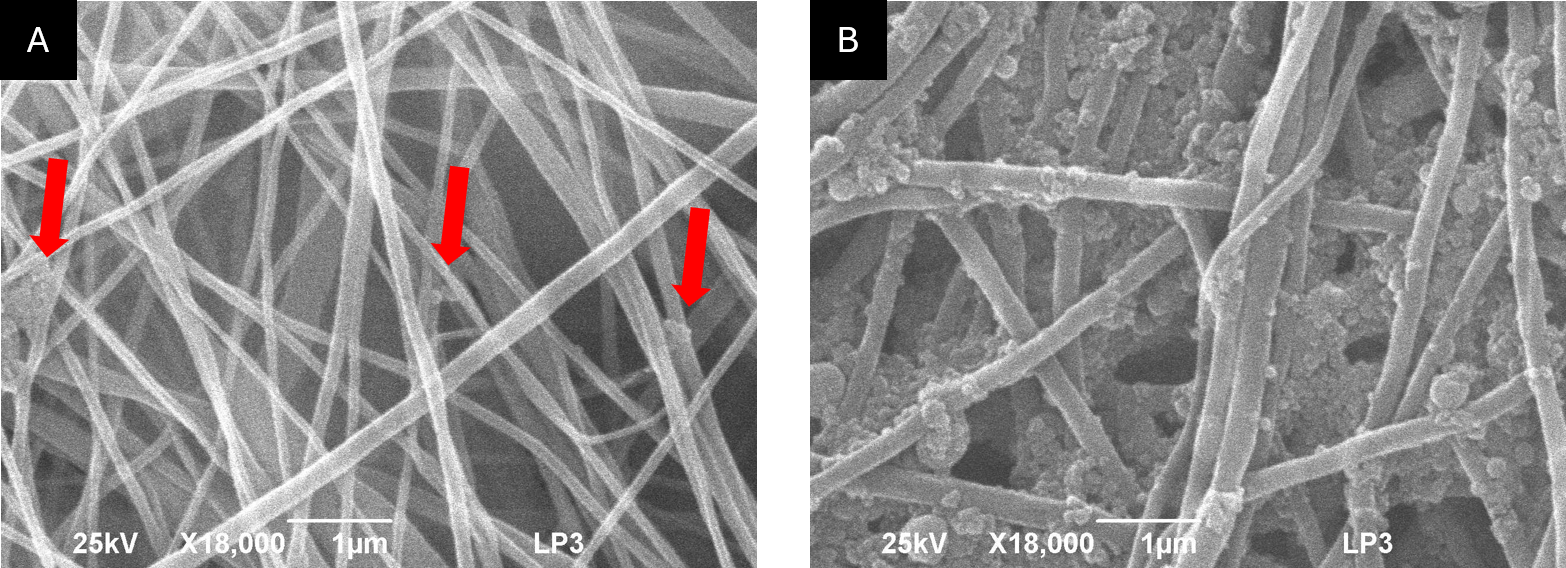}
\caption{Comparison between the ultrasonic deposition of SiNPs on PCL (A) and PA (B) samples. 
}\label{fig5}
\end{figure}

As seen in Fig.\ref{fig5}, SiNPs were not detected on the surface of the pure PCL samples submitted to immersion, but rather seemed aggregated in clusters that could occasionally be caught within the NF net. In contrast, PA samples showed numerous clusters of NPs all throughout the individual fibers surfaces, illustrating the successful grafting of the SiNPs and confirming the use of APTES as an adequate linker between the PCL NFs and the SiNPs. This phenomenon was attributed to the previously described unfavorable interactions between the PCL NFs and the suspended SiNPs due a lack of chemical affinity between the polymer and the nanoparticles. The rare remaining clusters of NPs detected seemed to be immobilized in a physical manner, wedged between fibers. Samples functionalized with APTES showed effective functionalization, which was attributed to the silane's ability to chemically bond with the surface of the SiNPs, as was described in the previous section. This result shows that not only is the APTES contained within the NFs, but it is also present in the surface, and the silane moiety was not degraded by the electrospinning process, allowing for post-synthesis functionalization. However, ultrasonication of the materials for long periods of time caused significant degradation of the aluminum support onto which the NFs were deposited, leaving aluminum flakes in the bottom of the beaker. Reducing the treatment time to 5 minutes was sufficient to avoid this issue and this duration was thus selected for the subsequent experiments.

\subsection{Mat surface rugosity and pore size measurements}
\label{subsec3.4}

All formulations prepared were then characterized further by studying their surface layers. Rugosity is critical in the understanding of the liquid-NF mat interaction. To this end, each sample's surface was scanned using a laser microscope at a 100x magnification, granting a high-definition roughness evaluation. The obtained results are reported in Table.\ref{tab1}.

Pure PCL nanofibers present the highest roughness, with an Ra value of 0.973$\pm$0.077 $\mu$m. This illustrates a relatively irregular surface, as is often found within electrospun mats. Indeed, the presence of larger nanofibers, inhomogeneities and the overall highly porous structure of the mats can lead to higher roughness values. Introducing APTES to the polymer melt leads to lower roughness values overall, going as low as 0.593$\pm$0.062 for the PAC-16 sample. 

APTES has effects on the polymer viscosity, chemical affinity of the material constituents and homogeneity of the electrospun membranes, as was described in previous sections. In term, a higher material uniformity leads to a smoother surface presenting less defects. Interestingly, Ra seems to evolve in a non-linear manner in the "volume" samples, with higher values reported for PAC1 and PAC4 when compared to PA. We have observed a similar phenomenon when previously evaluating the mechanical properties of the NF mats. It would seem that lower concentrations of SiNPs in the volume of the fibers can produce occasional clumps acting as defects in the membrane, a phenomenon well studied in the litterature \cite{zaccaria2012effect}. A higher concentration of NPs can counteract this effect, producing a homogenous network of NPs, distributing stress efficiently \cite{li2021role}. A similar effect could be at play here, with the higher concentration of NPs increasing the homogeneity of the fibers and thus leading to a smoother surface.  
When submitting these samples to ultrasonication however, roughness is significantly decreased following the concentration of SiNPs employed. As the use of ultrasounds in post-treatment can spread the NFs in a mat, increasing pore size, a side effect could be the increase in mat homogeneity and thus lower Ra \cite{lee2011highly}. The increasing concentration of NPs deposited on the surface could then coat the relaxed fibers, minimizing the irregularities of the material. 
 
\begin{figure}[htbp]
\centering
\includegraphics[scale=0.22]{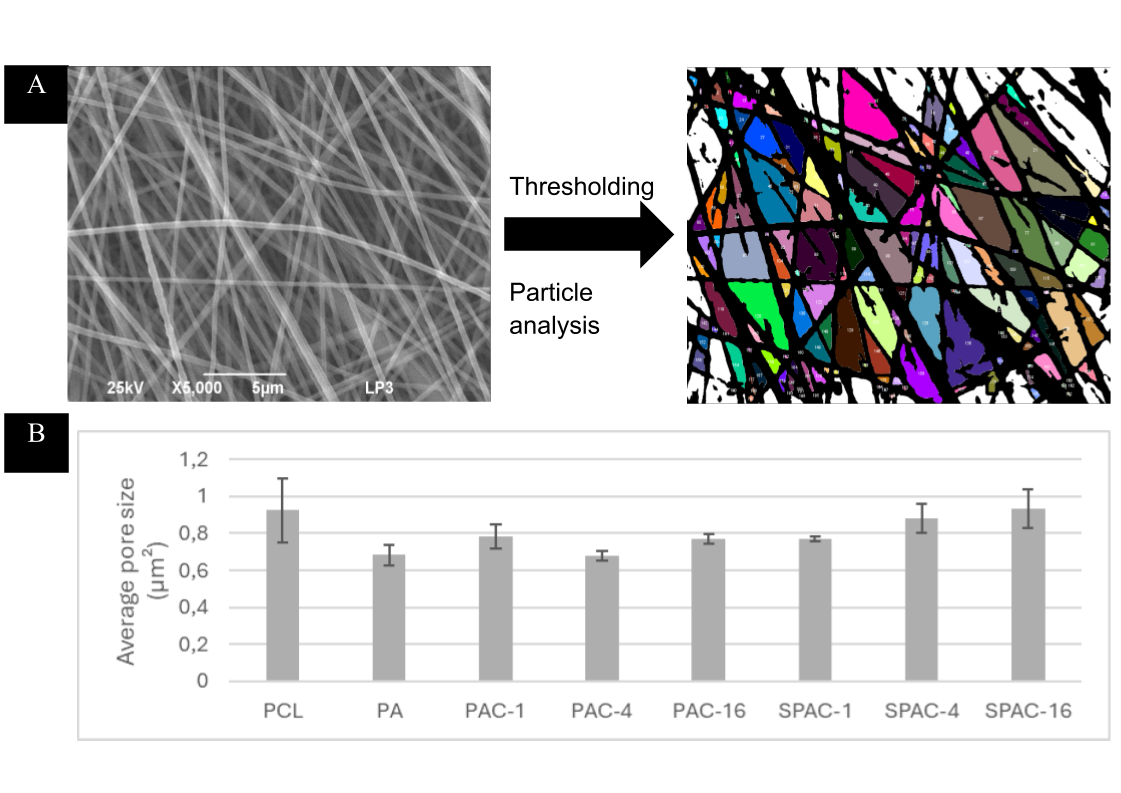}
\caption{Pore size measurements. (A) Image treatment illustration (B) Reported pore size values}\label{fig6}
\end{figure}

\begin{table}[htbp]
\begin{tabular}{|c|c|}
\hline
Sample & Mat roughness Ra ($\mu$m)\\
\hline
PCL  & 0.973 $\pm$ 0.077  \\
\hline
PA & 0.631 $\pm$ 0.046   \\
\hline
PAC-1  & 0.729 $\pm$ 0.081  \\
\hline
PAC-4 & 0.769 $\pm$ 0.049   \\
\hline
PAC-16  & 0.593 $\pm$ 0.062  \\
\hline
SPAC-1 & 0.528 $\pm$ 0.054   \\
\hline
SPAC-4  & 0.439 $\pm$ 0.019  \\
\hline
SPAC-16 & 0.362 $\pm$ 0.022   \\
\hline
\end{tabular}
\centering
\caption{Mat roughness data of different sample used in this study.}
\label{tab1}
\end{table}
Additionally, considering the highly porous structure inherent to the NF mats, the pore size of our samples was evaluated digitally from SEM micrographs (Fig.\ref{fig6}). Thresholding at 50$\%$ allowed us to isolate the upper layers of the NFs, as considering deeper layers might lead to increased overlap, introducing bias in the pore size measurements. Pore size remained between 0.6 and 1.0 $\mu$m$^{2}$ overall, with fluctuations seemingly dependent on the type of functionalization. Pure PCL NFs showed greater pore sizes around 0.925 $\pm$ 0.172 $\mu$m$^{2}$, a significant difference with the rest of the samples. In comparison, samples containing SiNPs in the volume of the fibers showed an average pore size of 0.742 $\pm$ 0.057 $\mu$m$^{2}$, and the samples functionalized on their surface presented pores of 0.861 $\pm$ 0.083 $\mu$m$^{2}$.  This discrepancy was attributed to multiple factors. Firstly, it was demonstrated that the viscosity of the electrospinning solutions diminished significantly when APTES was introduced. This could in term lead to thinner fibers and quicker material deposition, as the polymer ejection speed is greatly affected by its viscosity \cite{xue2024impacts}. Thus, samples containing APTES, with thinner fibers overall and more material deposited over the same electrospinning duration could present more layers of NFs, thus reducing the overall pore size. In contrast, submitting these NFs to ultrasounds when immersed in NP suspension increased the pore area. This effect has been described in literature, as the ultrasonication of NFs leads to fiber relaxation, increasing pore size. In some cases, this effect allowed for greater cell infiltration of the electrospun membranes \cite{lee2011highly, rafiq2023improvisations}. 

\subsection{Dynamic wetting of blood on functionalized PCL nanofibers}
\label{subsec3.5}

\begin{figure}[htbp]
\centering
\includegraphics[scale=0.22]{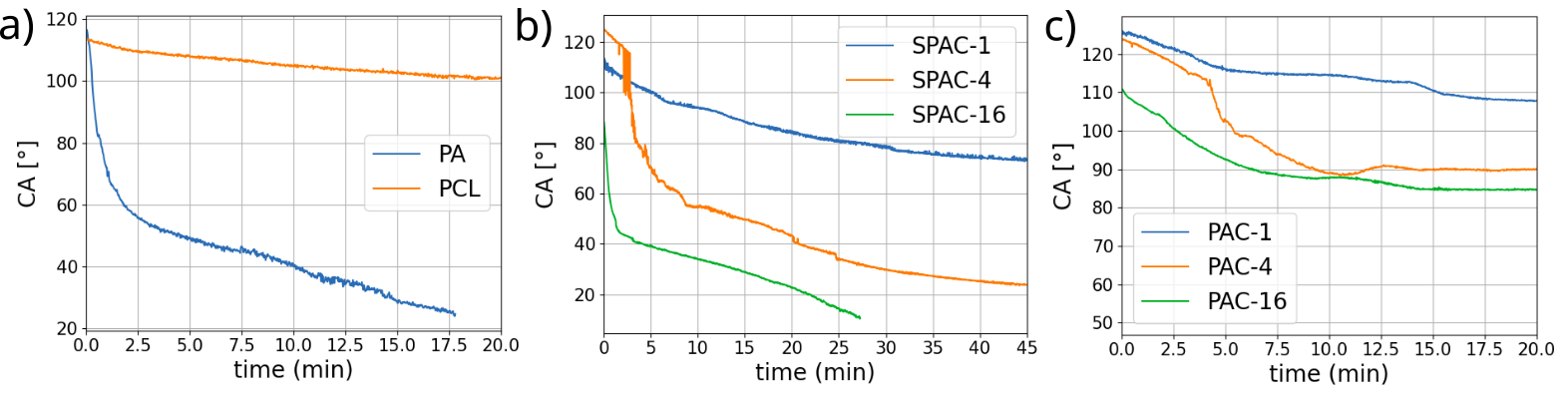}
\caption{Time evolution of blood contact angle measured on electrospun substrates. a) Pristine polycaprolactone (PCL) and PCL functionalized with APTES (PA). b) PCL nanofibers functionalized with laser-synthesized silicone nanoparticles incorporated within the fiber volume at different concentration (PAC1, PAC4, PAC16). c) PCL nanofibers functionalized with silicon nanoparticles deposited at the fiber surface at different concentration (CPAC1, CPAC4, CPAC16).}\label{fig7}
\end{figure}

Figure.\ref{fig7} summarizes the wetting dynamic behavior of whole blood on different substrates. In all cases the apparent contact angle (CA) decreases with time, reflecting spreading of the drop and progressive imbibition of the liquid into the fibrous network, but the rate and magnitude of these decreases are strongly dependent on the substrate chemistry and on the presence and localization of the Si nanoparticles.

On the reference substrates (Fig.\ref{fig7} a), pristine eletrospun PCL exhibits a high initial CA of about 110°, which only slowly decreases to about 100° over 20 minutes. This behavior confirms the well-known hydrophobic nature of PCL. In contrast, the PA substrates show a strong decrease of CA during time, the CA drops from 115° to 60° within the first 2-3 minutes and continues to decrease to a value below 30° at 20 minutes. The quick loss of hydrophobicity on the PA suggests a significantly higher surface energy and/or more efficient capillary absorption of blood in the porous structure compared to PCL. The final CA value indicates the complete wetting of blood. Those measurements showed that the electrospun PCL is considerably less wettable by blood than PA. PCL is a slow-degrading polymer already approved for medical application and widely used in blood-contacting devices, whereas PA is non-degradable and less relevant for long-term vascular application. Accordingly, the subsequent analysis focuses on PCL functionalized with laser-synthesized Si nanoparticles. The functionalization of the PCL will modify the interfacial chemistry -key parameters governing blood responses and protein adsorption- while preserving the bulk mechanical properties and degradation kinetics of the PCL matrix.

Figure.\ref{fig7}-b shows that all "volume" composites start with relatively CA values similar to pristine PCL (115°-125°), which indicate that the outer surface is still predominately composed of PCL. For the lowest concentration PAC1, the CA decreases slightly to about 112°, showing behavior similar to the pristine eletrospun PCL. Increasing the amount of Si nanoparticles in volume accelerates the CA decay; in the PAC4 sample the CA tends to 90° at 20min, and in the PAC16 sample it decreases further to 85°. This behavior can be attributed to several factors. The presence of rigid Si nanoparticles on the nanofiber can alter their morphology and packing, leading to variation in their diameter and higher surface area, which enhance capillary forces driving blood infiltration. Nevertheless, because most nanoparticles remain buried within the volume of fiber matrix, their contribution to the effective surface energy is limited, therefore, the substrates remain overall hydrophobic; with CA values higher than 85° even at high concentration of Si nanoparticles.

\begin{figure}[htbp]
\centering
\includegraphics[scale=0.52]{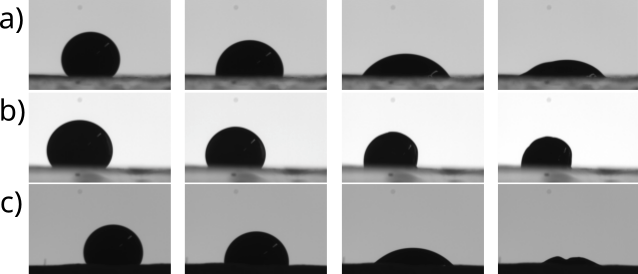}
\caption{Time evolution (from left to right) of blood droplet shapes on electrospun substrates. a) PCL functionalized with APTES (PA). b) PCL nanofibers functionalized with laser-synthesized silicone nanoparticles incorporated within the fiber volume at concentration C1. c) PCL nanofibers functionalized with silicon nanoparticles deposited at the fiber surface at concentration C16.}\label{fig8}
\end{figure}

When the Si nanoparticles are located at the surface of substrate, we observe a much stronger effect on blood wetting as shown in Fig.\ref{fig7}-c. For SPAC1, the CA decreases gradually, reaching 80° after 20 minutes, behavior similar to the case when the substrate is loaded in volume with high concentration of Si nanoparticles PAC16. On the other hand, the blood wetting on SPAC4 undergoes two-stage dynamics, a noticeable decline from 120° to about 60° within the first 5 minutes, followed by a slow decrease towards 20°, at the end of the experiment. A high surface loading of substrate with nanoparticles leads to a good hydrophilic response. The CA of blood droplet on SPAC16 drops from 100° to 40° in less than 2 minutes and continues to decrease linearly to reach a value close to 0 after about 25 minutes, indicating nearly complete wetting and efficient imbibition of blood throughout the fibrous network.

The comparison between Fig.\ref{fig7}-b and Fig.\ref{fig7}-c shows that not only the quantity of Si nanoparticles that control the blood wetting on substrate but also the spatial distribution of Si nanoparticles with the PCL scaffolds is important for controlling blood-material interactions. When the substrate surface is loaded with nanoparticles, their oxidized layers can interact directly with the liquid, creating a hydrophilic interface silanol-rich interface and an increase in surface free energy. Furthermore, hydrophilic behavior of the fibers promotes strong capillary suction in the inter-fiber pores, which enhances the transport of blood into the mat and explains the continuous decrease of CA over time. The pronounced dependence on nanoparticle concertation suggests that the shift from mainly hydrophobic behavior SPAC1 to highly wettable surface SPAC16 takes place within a relatively narrow window of surface coverage. From a practical standpoint, these results indicate that electrospun PCL by itself undergoes poor blood wettability, whereas functionalization with laser-synthesized silicon nanoparticles allows a robust and efficient means to tune the wetting dynamics response. By adjusting nanoparticle loading, the substrate evolves from near PCL-like behavior at low volume fraction to rapid and extensive spreading at high volume fraction, reaching values comparable to PA. This precise control over blood wettability is expected to directly impact the early stages of blood-material interactions, including plasma protein adsorption, conditioning layer formation, and, ultimately, the hemocompatibility of the nanofibrous scaffolds.

\begin{figure}[htbp]
\centering
\includegraphics[scale=0.45]{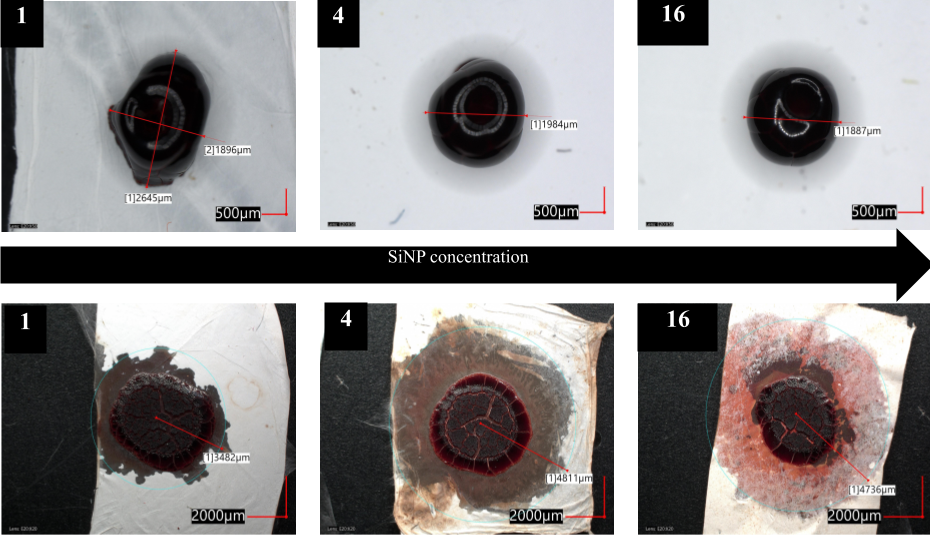}
\caption{Digital microscopy images of the blood droplets deposited on the various PCL NFs mats functionalized with SiNPs. The first row : samples with NPs in bulk. The second row : sample with NPs on surface. The value on each image indicates the concentration of NPs.}\label{fig9}
\end{figure}

Although dynamic contact angle measurements quantify the overall kinetics of spreading and imbibition, they provide only a one-dimensional outlook of blood-substrate interactions. They do not, by themselves, resolve how these wetting regimes shape the final spatial distribution of blood on and within the fibrous mats. To address this limitation, we therefore examined the three-dimensional morphology of the droplet at the final stage. 
Representative top-view of the dried blood drops on different PCL/SiNP mats are shown in Fig.\ref{fig9}. For the sample where SiNPs are confined in the fiber volume (upper row), the droplet remains compact and nearly hemispherical. The contours are sharp and there is only a narrow peripheral ring with slightly lighter color, indicating that a small quantity of plasma has spread into the surrounding mat. Blood mainly rests on the top of the hydrophobic PCL/SiNPs matrix and only weakly penetrates between nanofibers.

The situation is totally different when the nanoparticles are only loaded on the surface of the substrate (lower row). In this case, the central dark red region corresponding to the cell-rich fraction is surrounded by a broad, pale halo that extends a few millimeters from the drop center, with the total stained area increasing from SPAC1 to SPAC4. On SPAC4 and SPAC16, the central droplet appears concave, with irregular edges and a mottled outer zone, pointing to extensive wicking of the plasma phase into the porous scaffold and partial retention of red blood cells in the center. This morphology matches the strong decrease of contact angle and the large spreading/imbibition inferred from the dynamic measurements for the surface-functionalized mats, especially at the highest SiNPs loading. The digital micrographs thus offer a clear visual counterpart to the wetting measurement, showing that increasing the surface availability of Si nanoparticles progressively alters the blood-substrate interaction. The system evolves from a stable, sessile droplet on a predominately hydrophobic fibrous mat to a strongly infiltered imprint, in which plasma and cellular components redistribute and segregate within and around the nanofibrous matrix.

\begin{figure}[htbp]
\centering
\includegraphics[scale=0.65]{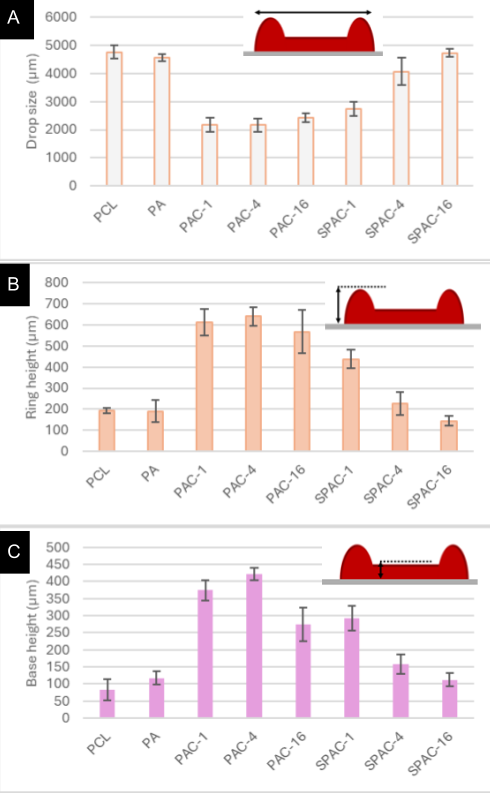}
\caption{Characterizations of the blood droplets deposited on the surface of the NF mats. (A) Drop diameter measurements; (B) Ring height measurements; (C) Drop base measurements}\label{fig10}
\end{figure}

Representative top‑view images of the dried blood drops on the different PCL/SiNP mats are shown in Fig.\ref{fig9}. For samples where SiNPs are embedded within the fiber volume (upper row of Fig.\ref{fig9}), the droplet remains compact and nearly hemispherical. The contour stay well defined, with only a thin peripheral ring with slightly lighter colour, indicating that only a small quantity of plasma penetrates inti the surrounding mat. The footprint diameter shows a slight reduction as the SiNP content increases (from C1 to C16), which is consistent with the relatively high contact angles and the modest time‑dependent decrease observed in the CA curves for the "volume" series. In this regime, blood remains on top of the hydrophobic PCL matrix, with only weakly spreading of plasma into the fibrous network.

The situation is totally different when the nanoparticles are only presented at the fiber surface (lower row of fig \ref{fig9}). In this case the central dark red region corresponding to the cell‑rich fraction is surrounded by a broad, pale halo that extends few millimeters from the droplet center, with the total stained area increasing from SPAC1 to SPAC4 and SPAC16. On SPAC4 and SPAC16, the contour of the central droplet is clearly identifiable, while a pronounced phase separation between plasma and red cells is observed. As the surface concentration of Si nanoparticles increases, the lateral extent of the plasma-rich region progressively expands beyond the red-cell-rich core, indicating enhanced spreading of the plasma phase along the nanofibrous surface. This behavior matches the strong decrease of contact angle and the large spreading/imbibition inferred from the dynamic measurements for the surface‑functionalized mats, especially at the highest SiNP loading. The digital micrographs therefore provide a direct visual counterpart of the wetting data, illustrating how increasing surface loading of SiNPs transforms the interaction with blood from a sessile droplet resting on a hydrophobic mat to a highly infiltrated stain where the different blood phases are spatially separated within and around the nanofibrous network.

\begin{figure}[htbp]
\centering
\includegraphics[scale=0.75]{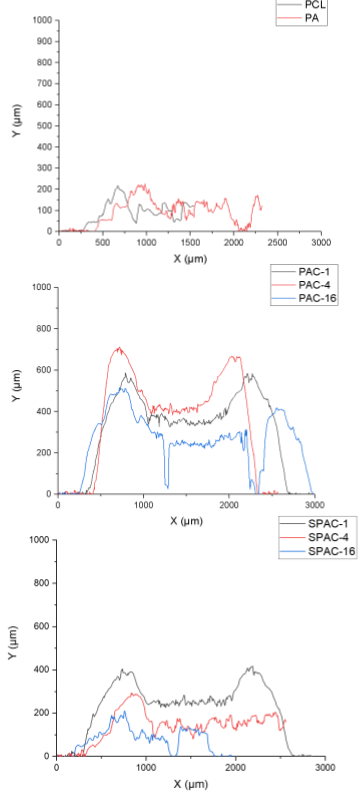}
\caption{Topological profiles of the blood droplets deposited on NFs, grouped by sample type.}\label{fig11}
\end{figure}

The three‑dimensional geometry of the dried blood droplets was next quantified by topographical microscopy in order to relate the dynamic contact‑angle behavior and the plan view images to the final deposit morphology. From the height maps we extracted the lateral drop diameter (Fig.\ref{fig10}A), the height of the peripheral ring (Fig.\ref{fig10}B) and the height of the central base (Fig.\ref{fig10}C), and representative cross‑sectional profiles are reported in Fig.\ref{fig11}.

On the reference mats (PCL and PA), the dried droplets display relatively large contact diameters ($\approx$ 4.5–5.0 mm) together with low ring and base heights ($\leq$ 250 µm), resulting in overall flattened profiles (Fig. \ref{fig10}B). The slightly larger diameter and thickness observed on PA are consistent with its more pronounced decrease in contact angle, suggesting that, despite substantial imbibition, a finite amount of blood remains distributed at the surface as a thin residual film.

A very different morphology is observed for the samples where SiNPs are confined within the PCL volume fibers (PAC‑1, ‑4, ‑16). Here the footprint diameter is significantly reduced ($\approx$ 3 mm), whereas both ring and base heights increase by a factor of three to four (up to $\approx$700 $\mu$m. The corresponding profiles (see Fig. \ref{11}) show narrow, tall caps with pronounced peripheral rims, characteristic of a droplet that has spread only marginally and dried mainly by evaporation rather than by wicking into the mat. This behavior is fully consistent with the high contact angles and limited CA decay measured for the "volume" composites.

When SiNPs are only presented at the fiber surface (SPAC‑1, ‑4, ‑16), the trend reverses. The drop diameter increases again and, for SPAC‑16, even exceeds that on the pristine matrices, while both ring and base heights progressively decrease with increasing SiNP surface coverage (Fig. \ref{fig11} A–C). The cross sections (Fig. \ref{fig11}) evolve from a peaked profile (SPAC‑1) to a broad and flattened shape for SPAC‑16, in agreement with the extensive lateral spreading and large stained areas observed in the top‑view images (Fig. \ref{fig9}). These behaviors indicate efficient capillary uptake of the plasma phase into the nanofibrous network and a redistribution of the cellular fraction, which attenuates the classical "coffee‑ring" pattern.

Taken together, the profilometry data corroborate the conclusions from the contact‑angle kinetics and digital microscopy: nanoparticles buried in the fiber volume have little effect on blood spreading, whereas SiNPs exposed at the fiber surface, particularly at the highest loading, fundamentally change the wetting regime from a compact sessile drop to a thin, highly infiltrated blood film. Such a morphology is expected to influence both the thickness and uniformity of the initial blood-derived layer that governs subsequent hemocompatibility.

\begin{figure}[htbp]
\centering
\includegraphics[scale=0.45]{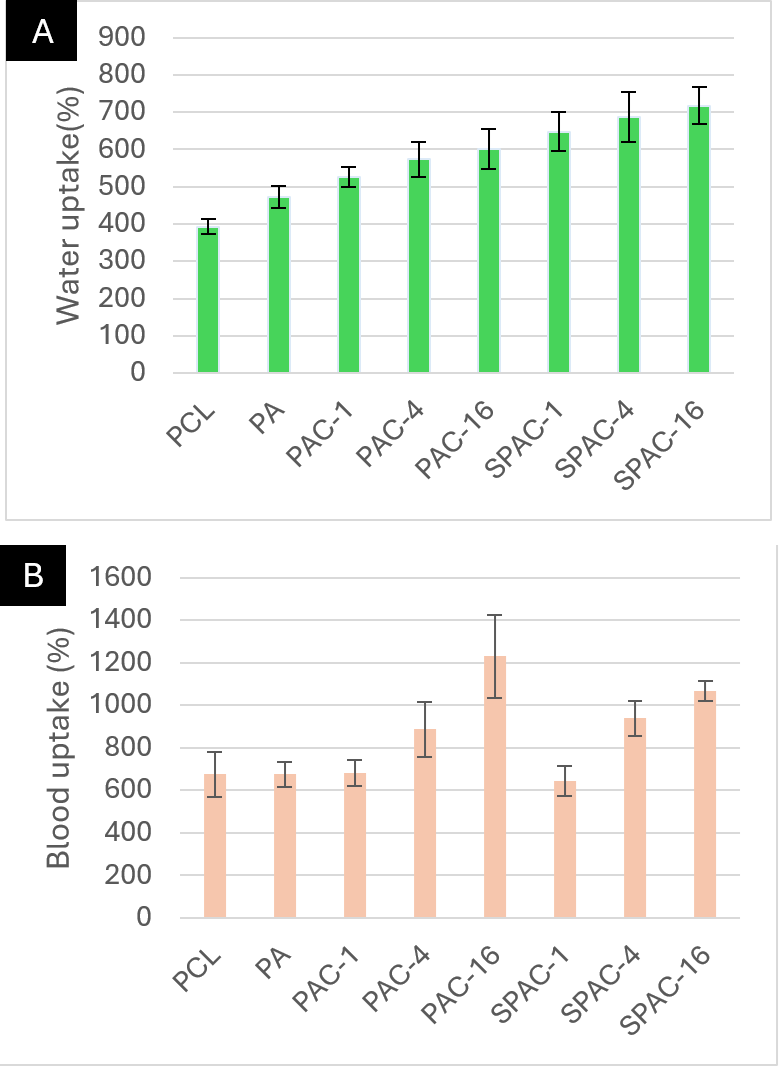}
\caption{Fluid absorption of the various NFs formulations. (A) Water uptake for samples immersed in deionized water for 1 minute; (B) Blood uptake for samples immersed in whole blood for 1 minute.}\label{fig12}
\end{figure}

The gravimetric fluid‐uptake measurements (Fig \ref{fig12}) complement the contact‐angle and droplet‑profilometry data by probing the capacity of the fibrous mats to absorb liquid throughout their thickness, under conditions where all faces of the scaffold are exposed to the medium. For deionized water (Fig \ref{fig12} A) the uptake of pristine PCL is the lowest ($\sim$400$\%$), whereas PA already shows a higher absorption ($\sim$480$\%$), in agreement with its stronger wetting kinetics. Incorporation of SiNPs into the PCL fibres (PAC series) leads to a progressive, almost linear increase in water absorption with nanoparticle content, reaching values close to 600$\%$ for PAC‑16. When the nanoparticles are preferentially located at the fiber surface (SPAC series), the water uptake is further enhanced and attains $\sim$700–750$\%$ for SPAC‑16.

Whole‑blood absorption (Fig \ref{fig12} B) follows a similar, although not identical, pattern. PCL, PA and the lowest‑loading formulations (PAC‑1 and SPAC‑1) display comparable uptake values around 650–700$\%$, suggesting that a modest increase in surface polarity is not sufficient to overcome the higher viscosity and cellular content of blood. At intermediate and high nanoparticle contents, however, blood uptake increases markedly. For PAC‑4 and SPAC‑4 the absorbed mass rises to $\sim$900–1000$\%$, and the highest values are found for PAC‑16, which exceeds 1200$\%$, with SPAC‑16 close to ($\sim$1050$\%$). These results are consistent with the top‑view and topographical observations: nanofibrous mats that permitted extensive droplet spreading and showed large stained areas (notably SPAC‑4 and SPAC‑16) also display enhanced macroscopic absorption, reflecting efficient capillary transport of the plasma phase through the network. The particularly high uptake of PAC‑16, despite its more compact sessile‑drop morphology, suggests that once blood penetrates the outer PCL layer, the presence of SiNPs throughout the fiber volume provides a dense network of hydrophilic sites that favors continued wicking from the bulk of the liquid. Overall, the fluid‑uptake data confirm that SiNP functionalization substantially increases the capacity of PCL NFs to accommodate both water and blood, with the exact magnitude depending on nanoparticle content and localization, in line with the wetting and droplet‑shape analyses.

\section{Conclusion}
\label{sect4}
This study has clarified how laser‑synthesized silicon nanoparticles (SiNPs) regulate the early interaction of whole blood with electrospun poly($\epsilon$‑caprolactone) (PCL) nanofibrous mats. By incorporating SiNPs either within the fiber bulk (PAC) or preferentially at the fiber surface (SPAC), we preserved the underlying fibrous architecture while systematically varying nanoscale interfacial chemistry. Dynamic contact angle measurements, together with digital microscopy and 3D profilometry, demonstrate that pristine PCL remains strongly hydrophobic, leading to tall and compact blood droplets. Embedding SiNPs within the fiber volume induces only a moderate increase in wettability and largely preserves this sessile-droplet morphology. In contrast, surface-decorated mats—most notably SPAC-16 display rapid contact-angle decay, extensive lateral spreading, markedly flattened profiles, and pronounced plasma–cell phase separation over extended surface areas.

Gravimetric measurements under immersion corroborated these findings: water and blood uptake increased with SiNP loading and were highest for SPAC‑16 and PAC‑16, demonstrating that SiNPs enhance not only surface wetting but also capillary imbibition through the fibrous network. The contrast between PAC and SPAC series highlights nanoparticle localization as a critical parameter: surface‑exposed SiNPs drive a transition from barrier‑like to highly absorbent behaviour, while particles in the fiber volume primarily affect bulk fluid uptake once the surface barrier is overcome.

These results emphasis that hemocompatibility of nanofibrous dressings is strongly influenced by early‑time wetting and blood transport, and that these processes can be tuned in a single polymer system without complex chemical grafting. For the colloid and interface science community, the PCL/SiNP system provides a useful model for studying non‑Newtonian wetting, phase separation and capillary flow in nanoparticle‑modified porous interfaces. Future work should couple time‑resolved imaging and modelling of blood infiltration with detailed analyses of protein adsorption and platelet response, to establish quantitative links between wetting regimes, interfacial structure of the adsorbed layer and subsequent biological outcomes.

\section{Ethical statement}

The project was approved by the agreement between partners CINaM and EFS (French Establishment for Blood Research).

\section{Declaration of competing interest}

The authors declare no competing interests.

\section{Acknowledgements}

The authors thank the DGA-AID and Aix-Marseille University for RS’ PhD funding. Authors thank the LaMP facilities at LP3 laboratory. This work was supported through the AAP-Margination by CNES (National Centre for Space Studies). M.A. was supported by the Turing Center for Living Systems (CENTURI), funded by France 2030, the French Government programme managed by the French National Research Agency (ANR-16-CONV-0001), and by the Excellence Initiative of Aix-Marseille University - AMIDEX.

\section{Data availability}

Data will be made available on request.

\bibliography{biblio}

\end{document}